\documentclass[%
reprint,
superscriptaddress,
nofootinbib,
 amsmath,amssymb,
 aps,
 prb
]{revtex4-2}
\usepackage[compat=1.1.0]{tikz-feynhand}
\usepackage{amsmath,amssymb,bm,braket,url,dsfont}
\usepackage{mathtools}
\usepackage{graphicx}
\usepackage{xcolor}
\usepackage{longtable}
\usepackage{multirow}
\usepackage{array}
\usepackage{booktabs}
\usepackage{physics}
\usepackage{here}
\usepackage[version=3]{mhchem}
\usepackage{siunitx}

\usepackage[
draft=false,
bookmarks=true,
bookmarksnumbered=false,
bookmarksopen=false,
bookmarksopenlevel=4,
colorlinks=true,
anchorcolor=black,
citecolor=blue,
filecolor=magenta,
linkcolor=blue,
linkbordercolor={1 0 0},
urlcolor=blue,
pdfborder={0 0 1},
pdftitle={},
pdfauthor={},
pdfsubject={},
pdfkeywords={},
pdfpagemode=UseThumbs,
]{hyperref}

\usepackage{orcidlink,ulem}

\newcommand{\expecval}[1]{\left \langle {#1} \right \rangle}
\newcommand{\figref}[1]{FIG.~\ref{#1}}
\newcommand{\tabref}[1]{TABLE~\ref{#1}}

\newcommand{\Eqref}[1]{Eq.~\eqref{#1}}

\begin{document}

\title{
        Effect of collective spin excitation on electronic transport in topological spin texture 
}
\author{Kohei Hattori}
\email{hattori-kohei053@g.ecc.u-tokyo.ac.jp}
\affiliation{Department of Applied Physics, The University of Tokyo, {Bunkyo}, Tokyo 113-8656, Japan}

\author{Hikaru Watanabe  \orcidlink{0000-0001-7329-9638}} 
\email{hikaru-watanabe@g.ecc.u-tokyo.ac.jp}
\affiliation{Research Center for Advanced Science and Technology, The University of Tokyo, {Meguro}, Tokyo 153-8904, Japan}

\author{Junta Iguchi}
\affiliation{Department of Applied Physics, The University of Tokyo, {Bunkyo}, Tokyo 113-8656, Japan}

\author{Takuya Nomoto  \orcidlink{0000-0002-4333-6773}} 
\affiliation{Research Center for Advanced Science and Technology, The University of Tokyo, {Meguro}, Tokyo 153-8904, Japan}
\affiliation{Department of Physics, Tokyo Metropolitan University, Hachioji, Tokyo 192-0397, Japan}

\author{Ryotaro Arita  \orcidlink{0000-0001-5725-072X}} 
\affiliation{Research Center for Advanced Science and Technology, The University of Tokyo, {Meguro}, Tokyo 153-8904, Japan}
\affiliation{Center for Emergent Matter Science, RIKEN, {Wako}, Saitama 351-0198, Japan}

\begin{abstract}
We develop an efficient real-time simulation method for the spin-charge coupled system in the velocity gauge.
This method enables us to compute the real-time simulation for the two-dimensional system with the complex spin texture.
We focus on the effect of the collective excitation of the localized spins on the electronic transport properties of the non-trivial topological state in real space.
To investigate this effect, we calculate the linear optical conductivity by calculating the real-time evolution of the Kondo lattice model on the triangular lattice, which hosts an all-in/all-out magnetic structure.
In the linear conductivity spectra, we observe multiple peaks below the bandgap regime, attributed to the resonant contributions of collective modes similar to the skyrmionic system, alongside broadband modifications resulting from off-resonant spin dynamics. 
The result shows that the collective excitation, similar to the skyrmionic system, influences the optical response of the electron systems based on symmetry analysis.
We elucidate the interference between the contributions from the different spin excitations to the optical conductivity in the multiple spin texture, pointing out the mode-dependent electrical activity.
We show the complex interplay between the complex spin texture and the itinerant electrons in the two-dimensional spin-charge coupled system.
\end{abstract}

\maketitle
\section{introduction}
Recent research intensively elucidated the effect of collective dynamics of the spontaneous order, such as magnetic order and excitonic order, on the optical responses of electronic systems~\cite{Sotome2021, Morimoto2016_exciton,Okamura2022, Iguchi,Murakami2020, Kaneko2021,Ono2021}.
For instance, theoretical studies simultaneously calculate the real-time evolution of the electron system and the order parameter~\cite{Iguchi,Murakami2020, Kaneko2021,Ono2021}.
More specifically, the study of magnetic order~\cite{Iguchi} shows that the collective excitation of the localized spins modulates the linear optical conductivity and photocurrent response in an antiferromagnetic chain.
The results identified the features which are absent within the independent particle approximation.
It is, however, not straightforward to apply the methodology to more complex systems, such as those in a two-dimensional and with a complex spin structure, due to their large computational cost.
The realization of efficient real-time simulation applicable to diverse cases is highly desirable since it is anticipated that one can explore intriguing spin-charge coupled dynamics, for example, by considering the magnetic order whose real-space topological texture may significantly influence electronic properties~\cite{Ohgushi2000, Batista2008,Nagaosa_2012, Nagaosa2013}.

The non-trivial topological state in real space appears in non-coplanar magnetic structures such as frustrated magnets and magnetic skyrmionic systems, where the complex spin texture formed by spin-orbit coupling~\cite{Muhlbauer} and itinerant-electron-mediated interactions \cite{Kurumaji2019,Hirschberger2019}.
Through the exchange interaction with localized spin systems having such nontrivial spin textures, the electron system feels the fictitious magnetic field and exhibits anomalous transverse transport.
Theoretical studies~\cite{Ohgushi2000, Batista2008,akagimotome2010, Hamamoto2015,Matsui2021, Feng2020, Zhou2016} have shown that the anomalous transverse conduction occurs in the Kondo lattice model with nontrivial spin structures.
Importantly, such non-coplanar spin textures, including the all-in/all-out (AIAO) structure \cite{Park2023,Takagi2023,Ghimire2018} and skyrmion crystal \cite{Kurumaji2019,Khanh2020,Hirschberger2019}, allows for collective modes richer than those in conventional ferro- and antiferromagnets.
Indeed, theoretical~\cite{Mochizuki2012,Mruczkiewicz2017,Mruczkiewicz2016} and experimental studies~\cite{Seki2020,Onose2012,Okamura2013} identified collective modes of the localized spins in the skyrmionic system. 
These collective modes include breathing modes excited by out-of-plane magnetic fields and clockwise and counterclockwise rotatory modes excited by in-plane magnetic fields~\cite{Mochizuki2012,Mruczkiewicz2017}.

In this study, we elaborate on the effect of the collective excitation of the AIAO spin texture on the anomalous transverse transport of the electron system. 
We simulate the real-time evolution of the Kondo lattice model on the triangular lattice hosting the AIAO spin texture comprised of four sites in the unit cell.
The dynamical property is clarified by computing the real-time evolution of the electrons and localized spins simultaneously. 
The computational cost of the two-dimensional spin-charge coupled system with the AIAO spin texture is expensive because of the dimensionality and complex spin structure. 
We resolve the difficulty by developing an efficient real-time simulation based on the velocity-gauge formulation and interpolation technique.
The developed calculation scheme allows us to demonstrate the intriguing interplay between the complex spin texture and itinerant electrons in a two-dimensional system.

The real-time simulation unambiguously shows that the optical conductivity spectrum of the electron system exhibits multiple peaks originating from rich collective dynamics of localized spins.
To classify which collective mode affects linear optical conductivity, we analyze the spin excitations by the symmetry-adapted bases.
The symmetry analysis indicates the similarity between the collective modes of the AIAO spin system and those of the skyrmionic system.
We find that the rotatory modes of the localized spins affect the optical responses of the electronic systems in two-dimensional systems.
Furthermore, the interference effect of the collective modes, which is characteristic of the system hosting the complex spin structure, results in the in-gap spectrum of the optical Hall conductivity whose intensity significantly differs between the modes.
These findings may lead to an understanding of the interplay between the complex spin texture and itinerant electrons.

The paper is organized as follows.
In Sec.~\ref{Sec_method}, we explain the details of the model and the computational scheme for the real-time evolution of the spin-charge coupled system.
Section~\ref{symmetry_analysis} is for the symmetry classification of the spin excitations and its coupling to the external stimuli.
Then, we elucidate the physical responses and corroborate the effects of the spin-charge coupled dynamics in Sec.~\ref{linear_response_functions}.
We draw the conclusion in Sec.~\ref{summary}.

\section{method}
\label{Sec_method}

\subsection{Model}

\begin{figure}
    \centering
    \includegraphics[width=\linewidth]{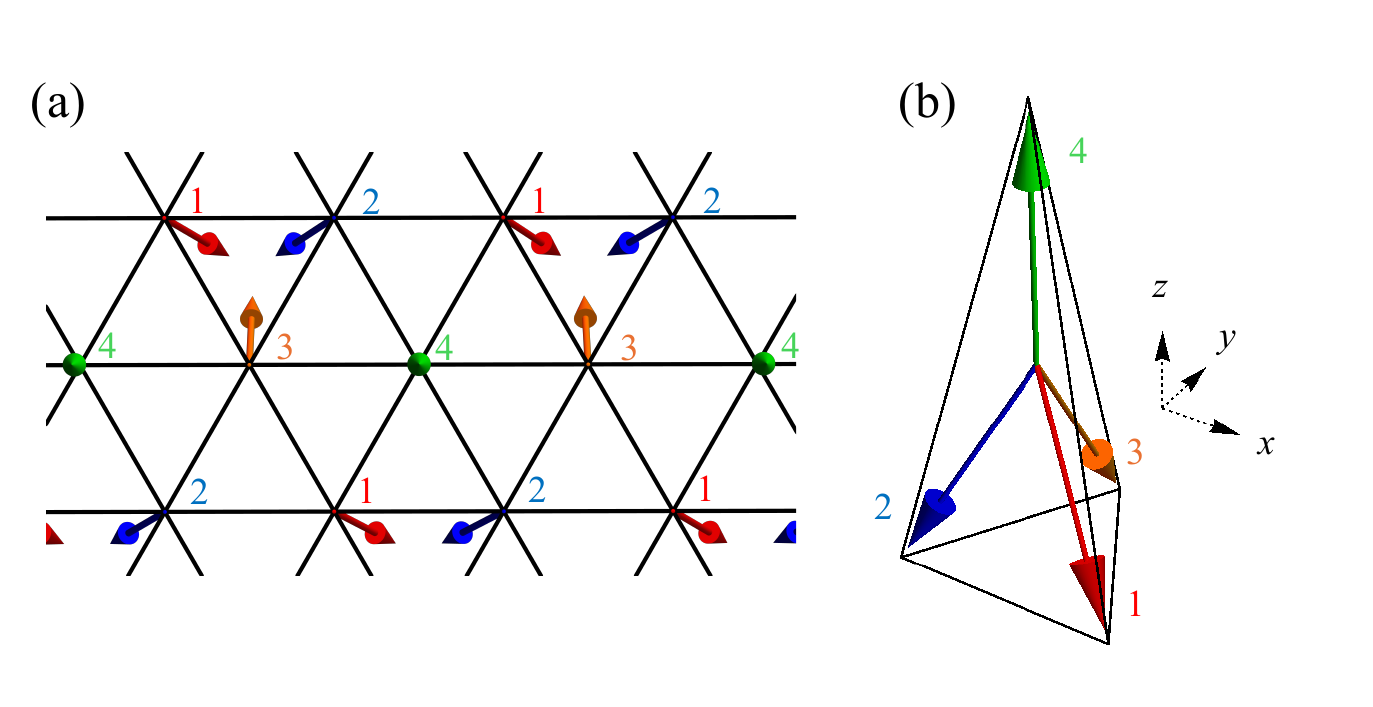}
    \caption{(a) Kondo lattice model with AIAO order on the triangular lattice with four-sublattice unit cell. (b) Spin moments at each sublattice. The spin texture is the AIAO state including canted component along the $z$-axis.}
    \label{model_pic}
\end{figure}

We work on the AIAO state in the Kondo lattice model on the triangular lattice (\figref{model_pic}). 
The tight-binding Hamiltonian is expressed as
\begin{align}
    \hat{\mathcal{H}} &= -\sum_{\langle ij\rangle\sigma\sigma'} t_{h} \hat{c}_{i\sigma}^{\dag} \hat{c}_{j\sigma'} - J \sum_{i\sigma\sigma'} \hat{c}_{i\sigma}^{\dag} \bm{\mathrm{S}}_i \cdot \bm{\sigma}_{\sigma\sigma'} \hat{c}_{i\sigma'} \nonumber \\
    &\quad -i\lambda \sum_{\langle ij\rangle\sigma\sigma'} \hat{c}_{i\sigma}^{\dag} \bm{e}_{ij} \cdot \bm{\sigma}_{\sigma\sigma'} \hat{c}_{j\sigma'} - K_z \sum_{i} ({\mathrm{S}}_i^z)^2.
    \label{Ham}
\end{align}
$\hat{c}_{i\sigma}^{\dagger}$ ($\hat{c}_{i\sigma}$) is the creation (annihilation) operator of the electron on the site $i$ with the spin index $\sigma$ ($\sigma=\uparrow, \downarrow$).
The first term of the Hamiltonian represents the hopping of the electrons between the nearest neighbor site $\langle ij\rangle$ with the amplitude $t_h$.
The second term represents the exchange interaction $J$ between the spin moment $\bm{\sigma}$ of the itinerant electron and the localized spin moment $\bm{\mathrm{S}}$.
The third term represents the antisymmetric spin-orbit coupling $\lambda$ (ASOC), which acts as an effective magnetic field parallel to the vector $\bm{e}_{ij}$ between the nearest neighboring sites $\langle ij\rangle$.
The ASOC breaks the inversion symmetry and thereby allows the system to exhibit the cross correlation between the spin and charge degrees of freedom.
The last term represents the easy anisotropy of the localized spin system $K_z$.

In the following calculations we use the parameters $t_{h} = 1,~J = 3.0,~\lambda = 0.3,~K_{z} = 0.1,~\alpha_{G} = 0.01,~\gamma = 0.01$, unless explicitly mentioned.
The filling number $n$ of electrons is set to $0.25$ per unit cell to stabilize the AIAO state in equilibrium [\figref{model_pic}~(b)]~\cite{akagimotome2010,Akagi2013}, where the system shows insulating behavior.
Owing to the anisotropy of the spin Hamiltonian, a finite magnetization appears along the $z$-axis.
We also set lattice constant $a = 1$ and elementary charge $e=1$.

We transform the Hamiltonian $\hat{\mathcal{H}}$ in the real-space representation into the momentum representation through the Fourier transformation
\begin{align}    
\hat{c}^{\dag}_{a\sigma}(\bm{k})=\sum_{\bm{R}}e^{i\bm{k}\cdot(\bm{r}_a+\bm{R})}\hat{c}^{\dag}_{a'\sigma},
\label{Bloch}
\end{align}
where the translation vectors $\bm{R}$ are defined as $\bm{R}=n_1(2,0,0)+n_2(1,\sqrt{3},0)$ $(n_1,n_2\in\mathbb{Z})$ and the position $\bm{r}_{a'}$ of the lattice site $a'$ is $\bm{r}_{a'}=\bm{r}_a+\bm{R}$.
Since the AIAO state has the unit cell four times larger than that in the paramagnetic state, the index $a$ for the sublattice in $\hat{c}^{\dag}_{a\sigma}(\bm{k})$ is taken as $a=1,2,3,4$ [\figref{model_pic}~(a)].
The Hamiltonian $\hat{\mathcal{H}}$ in the momentum representation reads as
\begin{align}
    \hat{\mathcal{H}}=\sum_{\bm{k}}\sum_{\sigma\sigma'}\sum_{ab}\qty(\bm{\mathcal{H}}_0(\bm{k}))^{\sigma\sigma'}_{ab}\hat{c}_{a\sigma}^{\dagger}(\bm{k})\hat{c}_{b\sigma'}(\bm{k}),
    \label{equilibrium-Hamiltonian-momentum-representation}
\end{align}
where $\bm{\mathcal{H}}_0(\bm{k})$ is the electronic Hamiltonian at each $k$-point.

When the electric field is incorporated under the velocity gauge, the creation (annihilation) operator changes as $\hat{c}_{i\sigma}^{\dag}\rightarrow e^{i\bm{A}(t)\cdot\bm{r}}\hat{c}_{i\sigma}^{\dag}\ (\hat{c}_{i\sigma}\rightarrow e^{-i\bm{A}(t)\cdot\bm{r}}\hat{c}_{i\sigma})$, where $\bm{A}(t)$ is the vector potential of the electromagnetic field.
In the velocity gauge, a time-dependent Hamiltonian is
\begin{align}
    \hat{\mathcal{H}}(t) = \sum_{\bm{k}}\sum_{\sigma\sigma'}\sum_{ab}\qty(\bm{\mathcal{H}}(\bm{k},t))^{\sigma\sigma'}_{ab}\hat{c}_{a\sigma}^{\dagger}(\bm{k})\hat{c}_{b\sigma'}(\bm{k}),
    \label{light-matter-coupling}
\end{align}
where the time-dependent electronic Hamiltonian $\bm{\mathcal{H}}(\bm{k},t)$ at each $k$-point is expressed as
\begin{align}
    \bm{\mathcal{H}}(\bm{k},t)=\bm{\mathcal{H}}_0(\bm{k}-\bm{A}(t)),
\end{align}
while the creation and annihilation operators are coincident with the original operators in \Eqref{equilibrium-Hamiltonian-momentum-representation} without the dependence on $t$.
The light-matter coupling is given by the time-dependent vector potential $\bm{A}(t)$ in the velocity gauge, which indicates the photo-electric field $\bm{E} (t) = -\partial_t \bm{A} (t)$.
Since the light-matter coupling is taken into account with the electric-dipole approximation, the vector potential is spatially uniform and thus does not break the translation symmetry.
In the expression \Eqref{light-matter-coupling}, we assume that the Wannier state of an electron is well-localized at a given site.

\subsection{Calculation Scheme}
In this study, we calculate the real-time evolution of the spin-charge coupled system.
To compute the real-time simulation, we simultaneously solve the von Neumann equation and the Landau-Lifshitz-Gilbert (LLG) equation. 

First, the time evolution of electrons can be described by the single-particle density matrix (SPDM) $\rho^{\sigma\sigma'}_{ab}(\bm{k},t) = \langle\hat{c}^{\dag}_{b\sigma'}(\bm{k})\hat{c}_{a\sigma}(\bm{k})\rangle_t\equiv\mathrm{Tr}[\hat{c}^{\dag}_{b\sigma'}(\bm{k})\hat{c}_{a\sigma}(\bm{k})\hat{\rho}(t)],$
where $\hat{\rho}(t)$ is the density matrix of the system at time $t$.
The SPDM satisfies the following equation called the von Neumann equation~\cite{Yue2022},
\begin{align}
    \frac{\partial\bm{\rho}(\bm{k},t)}{\partial t} = -i[\bm{\mathcal{H}}(\bm{k},t),\bm{\rho}(\bm{k},t)] - \gamma(\bm{\rho}(\bm{k},t) - \bm{\rho}_{\mathrm{eq}}(\bm{k},t)).
    \label{vonNeumann}
\end{align}
The vector potential $\bm{A}(t)$ for the light field does not break the translation symmetry, which keeps the von Neumann equation block-diagonal with respect to the momentum $\bm{k}$.
The $\bm{k}$-local property, which is compatible with the parallel computation, is advantageous for efficient simulation of the real-time evolution of spin-charge coupled dynamics.
On the other hand, if one works on the length gauge under which the light field is expressed by the scalar potential as Hamiltonian $- \bm{E}(t) \cdot \bm{r}$, the translation-symmetry breaking leads to $k$-derivative of the SPDM $-\bm{E}\cdot \frac{\partial\bm{\rho}(\bm{k},t)}{\partial \bm{k}}$ in the von Neumann equation.
This $k$-derivative term is unfavorable for parallel computation of $k$-point grids.

Second, the time evolution of the localized spin system is governed by the LLG equation
\begin{align}\label{LLG}
    \frac{d\bm{\mathrm{S}}_{a}}{dt} &= \frac{1}{1+\alpha_{\mathrm{G}}^2}\left(\bm{\mathrm{H}}_{a}^{\mathrm{eff}}\times\bm{\mathrm{S}}_{a} + \alpha_{\mathrm{G}}\bm{\mathrm{S}}_{a}\times(\bm{\mathrm{S}}_{a}\times\bm{\mathrm{H}}^{\mathrm{eff}}_{a})\right),\\
    \bm{\mathrm{H}}^{\mathrm{eff}}_{a} &= -J\langle\bm{\sigma}_{a}\rangle -2K_z (0,0,\mathrm{S}_a^z).
\end{align}
Here, $\bm{\mathrm{S}}_{a}$ represents the localized spin moment at sublattice site $a$, and $\bm{\mathrm{H}}_{a}^{\mathrm{eff}}$ represents the effective magnetic field coupled to $\bm{\mathrm{S}}_{a}$.
In the LLG equation \Eqref{LLG}, $\expecval{\vb*{\sigma}_{a}}$ is the sublattice-dependent spin density of electrons calculated with SPDM [see \Eqref{time-dependent-expectation-value}].
Furthermore, we account for a relaxation effect to obtain a physically reasonable response to light. 
Though relaxation stems from electron-electron correlations, electron-phonon interactions, and impurity scattering, we treat it phenomenologically by using the relaxation time approximation in the von Neumann equation as $\gamma(\vb*{\rho}(\bm{k}, t) - \vb*{\rho}_{\mathrm{eq}}(\bm{k},t))$ in \Eqref{vonNeumann}, and the Gilbert damping $\alpha_{G}$ in \Eqref{LLG}.
$\vb*{\rho}_{\text{eq}}(\bm{k},t)$ is the SPDM in equilibrium at the temperature $T=0$.
$\bm{\rho}_{\mathrm{eq}}(\bm{k},t)$ represents an equilibrium property but shows the dependence on time $t$ when the light-field is incorporated under the velocity gauge~\cite{Ventura2017,Murakami2022} (discussed below). 
We solve the coupled equations \Eqref{vonNeumann} and \Eqref{LLG}  by the fourth-order Runge-Kutta method.

In the velocity gauge, the time-dependent Hamiltonian indicates the modification of the velocity operator called the diamagnetic correction.
The correction effectively gives rise to the shift in momentum $\bm{k}$, and thus the momentum-resolved SPDM in equilibrium differs between those at initial time $t=0$ by the momentum shift.
The momentum shift explicitly reads as
\begin{align}
    \bm{\rho}_{\mathrm{eq}}(\bm{k},t) = \bm{\rho}^0_{\mathrm{eq}}(\bm{k}-\bm{A}(t)),
\end{align}
where $\bm{\rho}^0_{\mathrm{eq}}(\bm{k})$ is the SPDM for the initial Hamiltonian $\bm{\mathcal{H}}_0(\bm{k})$.
The SPDM for the equilibrium state, $\bm{\tilde{\rho}}_{\mathrm{eq}}(\bm{k},t)$ at time $t$, is calculated in the Bloch basis as follows
\begin{align}
    [\bm{\tilde{\rho}}_{\mathrm{eq}}(\bm{k},t)]_{mn} = \delta_{mn}\Theta(\mu - \epsilon_{n}(\bm{k},t)).
\end{align}
We can compute the SPDM for the equilibrium state in the orbital basis by using the unitary transformation
\begin{align}
    \bm{\rho}_{\mathrm{eq}}(\bm{k},t) = \bm{U}(\bm{k},t)\bm{\tilde{\rho}}_{\mathrm{eq}}(\bm{k},t)\bm{U}^{\dag}(\bm{k},t),
\end{align}
where the unitary matrix $\bm{U}(\bm{k},t)$ diagonalizes the Hamiltonian $\bm{\mathcal{H}}(\bm{k},t)$ as $\bm{U}^{\dag}(\bm{k},t)\bm{\mathcal{H}}(\bm{k},t)\bm{U}(\bm{k},t) = \bm{\mathcal{E}}(\bm{k},t)$ and $[\bm{\mathcal{E}}(\bm{k},t)]_{mn} = \delta_{mn}\epsilon_{n}(\bm{k},t)$.
The Hamiltonian changes as $\bm{\mathcal{H}}(\bm{k},t)=\bm{\mathcal{H}}_0(\bm{k}-\bm{A}(t))$ due to the electric field. Therefore, we need to calculate the SPDM $\bm{\rho}_{\mathrm{eq}}(\bm{k},t)$ at every time step by diagonalization of the Hamiltonian $\bm{\mathcal{H}}(\bm{k},t)$.

To avoid this large computational cost, we approximate the time-dependent SPDM in equilibrium.
We calculate the SPDM $\bm{\rho}_{\mathrm{eq}}(\bm{k},t)$ by applying cubic interpolation of the momentum grid.
We initially calculate the SPDM $\bm{\rho}^0_{\mathrm{eq}}(\bm{k})$ at $t=0$ and obtain the interpolated function $\bm{\rho}^{\mathrm{int}}_{\mathrm{eq}}(\bm{k})$ by performing the cubic interpolation of $\bm{\rho}^0_{\mathrm{eq}}(\bm{k})$ with respect to $\bm{k}$. 
Then we can approximate the SPDM $\bm{\rho}_{\mathrm{eq}}(\bm{k},t)$ at the time $t$ as
\begin{align}
    \bm{\rho}_{\mathrm{eq}}(\bm{k},t) \simeq \bm{\rho}^{\mathrm{int}}_{\mathrm{eq}}(\bm{k} - \bm{A}(t)).
\end{align}
By this interpolation technique, we can calculate $\bm{\rho}_{\mathrm{eq}}(\bm{k},t)$ easily without diagonalization of the Hamiltonian $\bm{\mathcal{H}}(\bm{k},t)$ at each time step.

Next, we explain a calculation method for physical responses.
At each time step, we evaluate the physical quantity $\langle\hat{\mathcal{O}}(t)\rangle$ as
\begin{align}
    \langle\hat{\mathcal{O}}(t)\rangle = \frac{1}{N}\sum_{\bm{k}}\mathrm{Tr}[\bm{\rho}(\bm{k},t)\bm{\mathcal{O}}(\bm{k},t)],
    \label{time-dependent-expectation-value}
\end{align}
by using the SPDM.
The current density operator is written by
\begin{align}
    \hat{\bm{j}}(t) &= \frac{1}{n_{\mathrm{sub}}}\sum_{\bm{k}}\sum_{ab}\sum_{\sigma\sigma^{\prime}}\pdv{\qty(\vb*{\mathcal{H}}(\bm{k},t))_{ab}^{\sigma\sigma^{\prime}}}{\bm{k}}\hat{c}_{a\sigma}^{\dagger}(\bm{k})\hat{c}_{b\sigma^{\prime}}(\bm{k}),\nonumber \\
        &\equiv \frac{1}{n_{\mathrm{sub}}}\sum_{\bm{k}}\sum_{ab}\sum_{\sigma\sigma^{\prime}}\qty(\vb*{j}(\bm{k},t))_{ab}^{\sigma\sigma^{\prime}}\hat{c}_{a\sigma}^{\dagger}(\bm{k})\hat{c}_{b\sigma^{\prime}}(\bm{k}),
        \label{electric-current}
\end{align}
where $n_{\mathrm{sub}}$ represents the number of sublattices in a unit cell.
Then, the linear response function is obtained as follows.
We apply an external field with the Gaussian profile described as
\begin{align}\label{Gauss}
    \bm{F}(t) &= \frac{\bm{F}_0}{\sqrt{2\pi\sigma^2}}\mathrm{exp}\left(-\frac{(t-t_0)^2}{2\sigma^2}\right).
\end{align}
Since we took the velocity gauge by which $\bm{E}(t) = -\frac{\partial\bm{A}(t)}{\partial t}$ holds, the Gaussian electric field $\bm{E}(t)$ is incorporated by the vector potential $\bm{A}(t)$ given by
\begin{align}
    \bm{A}(t) = \frac{\bm{E}_0}{2}\left[\mathrm{erf}\left(-\frac{t_0}{\sqrt{2}\sigma}\right)-\mathrm{erf}\left(\frac{t-t_0}{\sqrt{2}\sigma}\right)\right],
\end{align}
with the error function $\mathrm{erf} (\cdot )$.
In this scheme, we can calculate the linear response function $\chi_{\mathcal{O}F}(\omega)$ of the physical quantity $\mathcal{O}(t)$ to the external field $F(t)$ by the Fourier transformation
\begin{align}
    \chi_{\mathcal{O}F}(\omega) &= \frac{\Delta\mathcal{O}(\omega)}{F(\omega)},\nonumber\\
    &= \frac{1}{F_0}e^{\frac{\sigma^2\omega^2}{2}}e^{i\omega t_0}\int_0^{\infty} \mathcal{O}(t)e^{-i\omega t}dt,
    \label{Fourier}
\end{align}
where $\Delta\mathcal{O}(\omega)$ is the Fourier component of $\Delta\mathcal{O}(t)=\mathcal{O}(t)-\mathcal{O}(0)$ and $F(\omega)$ is the Fourier component of the external field $F(t)=\frac{F_0}{\sqrt{2\pi\sigma^2}}\mathrm{exp}\left(-\frac{(t-t_0)^2}{2\sigma^2}\right)$.
In this calculation, we use the parameter of the external field $F_0=1.0\times10^{-5},~t_0=0.2,~\sigma=0.03$.
Based on this scheme, we calculate the linear response functions to the light field, which will be shown in the next section.

We set the $k$-mesh of the Brillouin zone to $512\times512$ for the real-time simulation and the $k$-mesh of the Brillouin zone to $1600\times1600$ for the cubic interpolation of the SPDM in equilibrium.
In this calculation, we implement the parallel computation for $k$-mesh by the message passing interface (MPI).

\section{Symmetry Analysis}\label{symmetry_analysis}

\begin{table*}
        \caption{Eigenvalue of symmetry-adapted bases of localized spins under a symmetry operation in magnetic point group $\mathcal{G}$. The subscript $-$ of $\psi_\alpha$ represents the clockwise (CW) rotation, and the index $+$ represents the counterclockwise (CCW). In the columns of $3_z$ and $\theta 2_x$, $+1$ means the basis does not change its sign, while $-1$ means that the operation flips the sign. $\xi^-=\mathrm{exp}(2\pi i/3)$ denotes the eigenvalue of the CW mode and $\xi^+=\mathrm{exp}(-2\pi i/3)$ is that of the CCW mode.
        Each basis function $\psi_\alpha$ is explicitly written in the column `Basis function'.
        }
        \begin{ruledtabular}
        \begin{tabular}{cccc}
         $\psi_\alpha$       & $3_z$ & $\theta2_{x}$&Basis function  \\\colrule
        $\mathrm{A}_1^{1z}$&  $+1$           & $+1$ &$\mathrm{S}_4^z$        \\
        $\mathrm{A}_1^{3z}$&  $+1$           & $+1$  &$\frac{1}{\sqrt{3}}(\mathrm{S}_1^{z}+\mathrm{S}_2^{z}+\mathrm{S}_3^{z})$      \\
        $\mathrm{A}_1^{3xy}$ & $+1$           & $+1$ &$\frac{1}{\sqrt{3}}\left[(\frac{\sqrt{3}}{2}\mathrm{S}_1^x-\frac{1}{2}\mathrm{S}_1^{y})+(-\frac{\sqrt{3}}{2}\mathrm{S}_2^x-\frac{1}{2}\mathrm{S}_2^{y})+\mathrm{S}_3^y\right]$       \\
        $\mathrm{A}_2^{3xy}$ & $+1$           & $-1$&$\frac{1}{\sqrt{3}}\left[(-\frac{1}{2}\mathrm{S}_1^x-\frac{\sqrt{3}}{2}\mathrm{S}_1^{y})+(-\frac{1}{2}\mathrm{S}_2^x+\frac{\sqrt{3}}{2}\mathrm{S}_2^{y})+\mathrm{S}_3^x\right]$        \\
        $\mathrm{E}_{\pm}^{1xy}$& $\xi^{\pm}$           & $-1$ &$\frac{1}{\sqrt{2}}(\mathrm{S}_4^x\pm i\mathrm{S}_4^y)$        \\
        $\mathrm{E}_{\pm}^{3z}$ & $\xi^{\pm}$           & $+1$ &$\frac{1}{\sqrt{2}}\left[\left(-\frac{1}{\sqrt{6}}\mathrm{S}_1^z-\frac{1}{\sqrt{6}}\mathrm{S}_2^z+\frac{2}{\sqrt{6}}\mathrm{S}_3^z\right)\pm i\left(-\frac{1}{\sqrt{2}}\mathrm{S}_1^z+\frac{1}{\sqrt{2}}\mathrm{S}_2^z\right)\right]$        \\
        $\mathrm{E}_{\pm}^{3xy}$ & $\xi^{\pm}$           & $+1$&$\frac{1}{\sqrt{2}}\left[\frac{1}{\sqrt{3}}\left\{(-\frac{\sqrt{3}}{2}\mathrm{S}_1^x-\frac{1}{2}\mathrm{S}_1^y)+(\frac{\sqrt{3}}{2}\mathrm{S}_2^x-\frac{1}{2}\mathrm{S}_2^y)+\mathrm{S}_3^y\right\}\pm\frac{i}{\sqrt{3}}\left\{(-\frac{1}{2}\mathrm{S}_1^x+\frac{\sqrt{3}}{2}\mathrm{S}_1^{y})+(-\frac{1}{2}\mathrm{S}_2^x-\frac{\sqrt{3}}{2}\mathrm{S}_2^{y})+\mathrm{S}_3^x\right\}\right]$        \\
        $\mathrm{E}_{\pm}^{3xy'}$ & $\xi^{\pm}$           & $-1$&$\frac{1}{\sqrt{2}}\left[\frac{1}{\sqrt{3}}(\mathrm{S}_1^x+\mathrm{S}_2^x+\mathrm{S}_3^x)\pm\frac{i}{\sqrt{3}}(\mathrm{S}_1^y+\mathrm{S}_2^y+\mathrm{S}_3^y)\right]$        \\
        \end{tabular}
        \label{symmetry-adapted}
        \end{ruledtabular}
\end{table*}

In this section, we present the symmetry analysis of the localized spin system.
First, we classify the collective modes of the localized spins.
Second, we analyze the collective spin dynamics that are linearly coupled to the external field based on the symmetry analysis.

\subsection{Symmetry-adapted basis of localized spins}

In this subsection, we analyze the collective modes of the localized spins in the framework of the magnetic representation and decompose them into three modes: the azimuthal symmetric mode (AS), clockwise mode (CW), and counterclockwise mode (CCW).

The system belongs to the magnetic point group $\mathcal{G}=32'$ and has the following symmetry
\begin{align}
    \mathcal{G}=\{1,3_z,3_z^{-1},\theta 2_x,\theta 2_u,\theta 2_v\}.
\end{align}
Here, $1,\theta,3_a,2_a$ are the identity operator, time-reversal operator, threefold rotation around the $a$ axis, and twofold rotation around the $a$ axis, respectively. 
The $u$ and $v$ axes are generated by the action $3_z,3_z^{-1}$ on the $x$ axis, respectively.
The representation $\Gamma$ is obtained for the basis spanned by the degrees of freedom about localized spin 
vectors $\mathrm{S}_a^{\mu}$.
The dimension of $\Gamma$ is 12 due to the sublattice $(a = 1,2,3,4)$ and spins $(\mu = x,y,z)$.
The representation is decomposed into the irreducible representations as
\begin{align*}
    \Gamma=3\mathrm{A}_1\oplus\mathrm{A}_2\oplus4\mathrm{E}.
\end{align*}
There are three symmetry-adapted bases for the irreducible representation $\mathrm{A}_1$, one for $\mathrm{A}_2$, and eight for $\mathrm{E}$.

We further classify the symmetry-adapted modes by using the three-fold rotations.
We define the AS mode by the mode with its eigenvalue $1$ under the symmetry operation $3_z$, the CW mode as that with $\xi^{-}=\mathrm{exp}(2\pi i/3)$, and the CCW mode as that with $\xi^+=\mathrm{exp}(-2\pi i/3)$.
The symmetry-adapted basis function is given by the linear combination of the localized spins $\mathrm{S}_{a}^{\mu}$ as
\begin{align}
    \psi_{\alpha}=\sum_{a\mu}c^{\mu}_{a}(\alpha)\mathrm{S}_{a}^{\mu},
\end{align}
where $c_a^{\mu}(\alpha)$ is the coefficient of the symmetry-adapted basis $\psi_\alpha$.

The dynamics of the localized spin system can be described by the symmetry-adapted basis listed in \tabref{symmetry-adapted}.
We also display the sketch of the symmetry-adapted basis functions $\left\{ \psi_\alpha \right\}$ in \figref{symm-ext}.
The AS bases, $\mathrm{A}_1^{1z}$, $\mathrm{A}_1^{3z}$, $\mathrm{A}_1^{3xy}$, and $\mathrm{A}_2^{3xy}$, are invariant under the out-of-plane rotation and thus are totally symmetric in the azimuthal plane.
The modes for clockwise rotation include $\mathrm{E}_{-}^{1xy}$, $\mathrm{E}_{-}^{3z}$, $\mathrm{E}_{-}^{3xy}$, and $\mathrm{E}_{-}^{3xy'}$, while the CCW mode comprises $\mathrm{E}_{+}^{1xy}$, $\mathrm{E}_{+}^{3z}$, $\mathrm{E}_{+}^{3xy}$, and $\mathrm{E}_{+}^{3xy'}$.
Thus, CW and CCW modes carry the nontrivial transformation property under the out-of-plane rotation.
Note that the CCW basis with $\xi_+=\mathrm{exp}(-2\pi i/3)$ is the complex conjugate of the CCW counterpart with $\xi_-=\mathrm{exp}(2\pi i/3)$.
For example, $\mathrm{E}_{+}^{1xy}$ is the complex conjugate of $\mathrm{E}_{-}^{1xy}$.
These collective modes are similar to those of the skyrmionic system \cite{Mochizuki2012,Mruczkiewicz2017}.

In the following sections, we calculate the component of the symmetry-adapted basis of the localized spin dynamics $\tilde{\psi}_{\alpha}(t)$ as
\begin{align}
    \label{inner_product}
    \tilde{\psi}_{\alpha}(t)=\sum_{a\mu}\left(c^{\mu}_{a}(\alpha)\right)^*\mathrm{S}_{a}^{\mu}(t),
\end{align}
where we project the localized spins $\bm{\mathrm{S}}_{a}^{\mu}(t)$ into the symmetry-adapted basis $\psi_{\alpha}$.

\subsection{Selection rule for optical excitation of spin dynamics}
\label{selection-rule-response}

\begin{table}
        \caption{Eigenvalue of external field of localized spins under symmetry operator of magnetic point group $\mathcal{G}$. The symmetry-adapted bases listed in the column of the basis function are linearly coupled to the external field in the row.}
        \begin{ruledtabular}
        \begin{tabular}{cccc}
                & $3_z$ & $\theta2_{x}$&Basis function \\\colrule
         $E^{\pm}$ &  $\xi^{\pm}$           & $-1$&$\mathrm{E}_{\pm}^{1xy},\mathrm{E}_{\pm}^{3z},\mathrm{E}_{\pm}^{3xy},\mathrm{E}_{\pm}^{3xy'}$       \\
                $B^{\pm}$ &  $\xi^{\pm}$           & $+1$ &$\mathrm{E}_{\pm}^{1xy},\mathrm{E}_{\pm}^{3z},\mathrm{E}_{\pm}^{3xy},\mathrm{E}_{\pm}^{3xy'}$          \\
                $B^{z}$ &  $+1$           & $+1$ &$\mathrm{A}_{1}^{1z},\mathrm{A}_{1}^{3z},\mathrm{A}_{1}^{3xy},\mathrm{A}_{2}^{3xy}$
        \end{tabular}
        \label{symm-ext}
        \end{ruledtabular}
\end{table}

\begin{figure*}
        \centering
        \includegraphics[width=\linewidth]{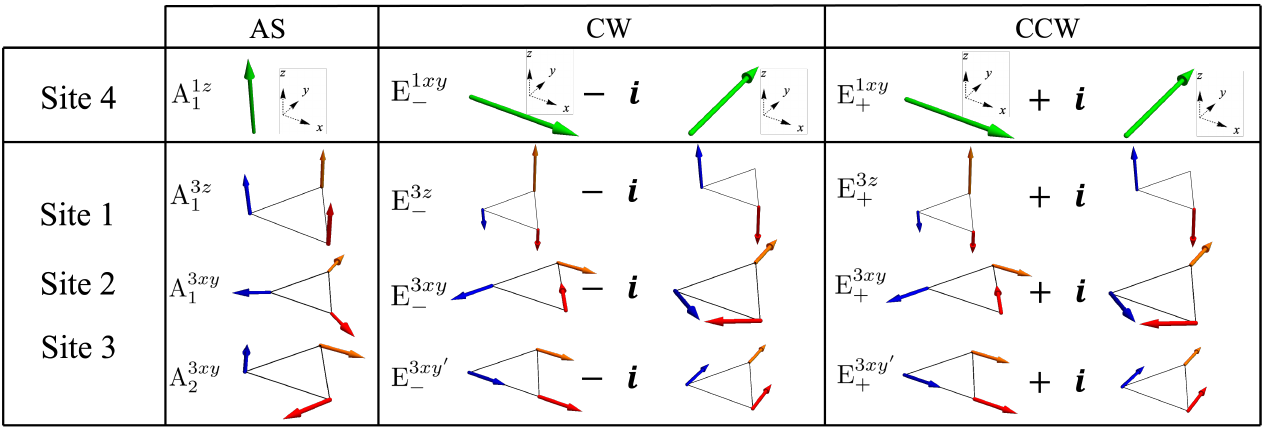}
        \caption{
        Sketch of the symmetry-adapted basis of localized spins tabulated in \tabref{symm-ext}.
        The spin moments are colored as done in \figref{model_pic}. 
        }
        \label{pic_basis}
\end{figure*}

In this subsection, we analyze the symmetry-adapted bases of the localized spins that are linearly coupled to the external field based on the symmetry.
The circularly polarized external field is defined as
\begin{align*}
    \bm{F}_{\pm}(t)= \frac{1}{\sqrt{2}}(1, \pm i, 0)\, \mathrm{exp}(i\omega t),
\end{align*}
where the minus sign ($-$) represents the CW mode and the plus ($+$) represents the CCW mode.
Similarly to \tabref{symmetry-adapted}, the external fields are classified in terms of irreducible representations and eigenvalues for the operation $3_z$. 
The result is shown in \tabref{symm-ext}.
For instance, the eigenvalue of the clockwise rotatory field $\bm{F}_{-}(t)$ is $\xi^{-}$ for the operation $3_z$, equal to that of the CW mode.

The symmetry indicates that we can excite the CW mode by the clockwise rotatory field $\bm{F}_{-}(t)$ and the CCW mode by the counterclockwise rotatory field $\bm{F}_+(t)$.
The linearly polarized field $F_x$ contains the clockwise and the counterclockwise rotation fields and can excite both the CW and CCW modes.
On the other hand, the eigenvalue of the magnetic field $B_z$ is $1$ for $3_z$, which is the same as that of the AS mode.
Thus, we can excite the AS mode by the magnetic field $B_z$ along the $z$ axis.
In the following parts, we mainly delve into the linear responses to in-plane electric fields, and the AS modes play minor roles.

\section{Linear Response Functions }\label{linear_response_functions}

In this section, we present the results of the real-time simulation of the spin-charge coupled system.
First, we calculate the magnetic susceptibility of the symmetry-adapted bases of the localized spin system in Sec.~\ref{mag}.
Second, we calculate the linear optical conductivity of the electron system in Sec.~\ref{opt}.
Third, we calculate the electromagnetic susceptibility of the localized spin system in Sec.~\ref{ele}.
We also quantitatively evaluate the electromagnetic susceptibility of the collective modes.

\subsection{Magnetic Susceptibility}\label{mag}

\begin{figure}
        \centering
        \includegraphics[width=\linewidth]{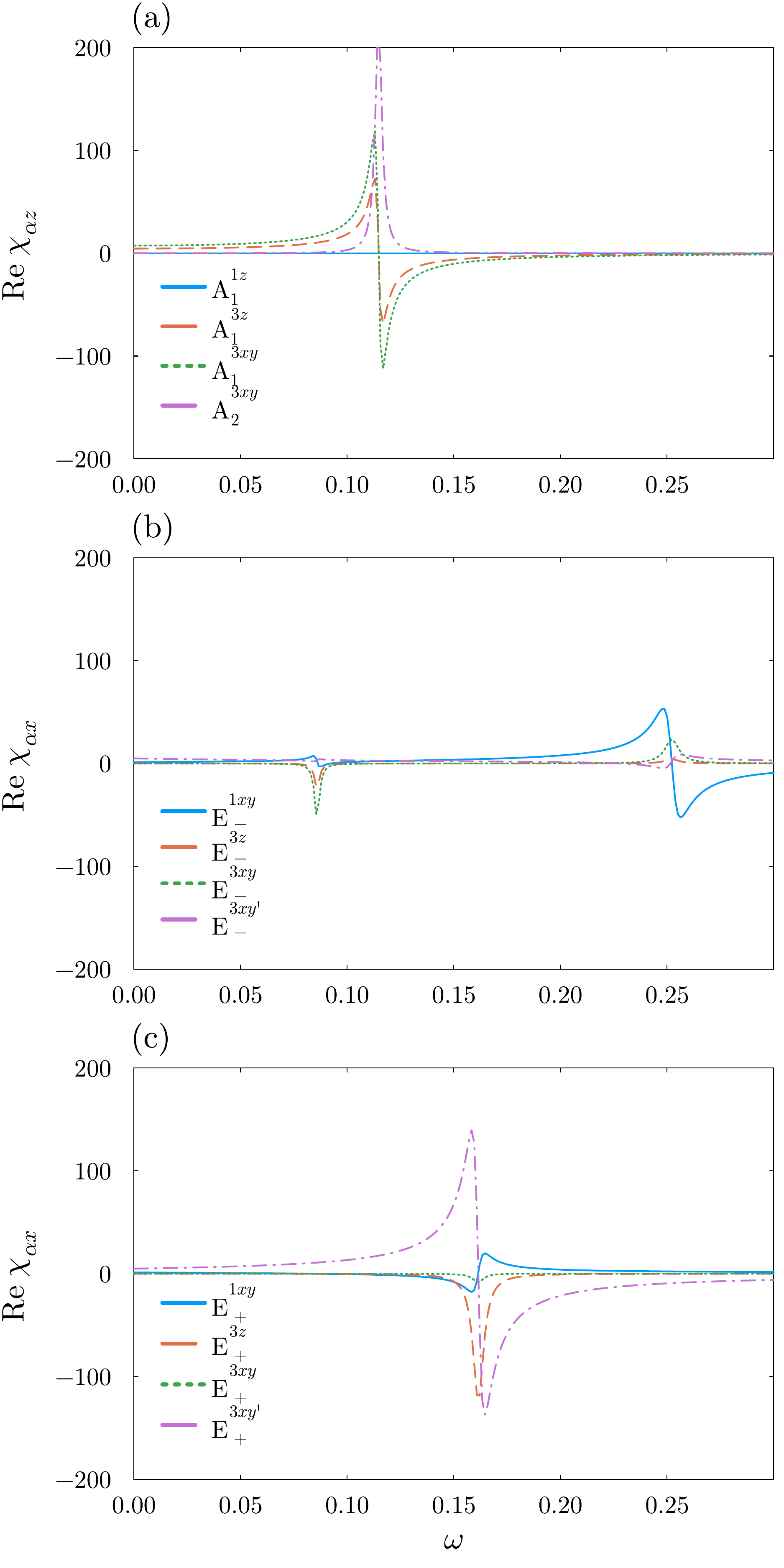}
        \caption{(a) Linear magnetic susceptibility of the AS mode to the out-of-plane magnetic field. (b) Linear magnetic susceptibility of the CW mode to the magnetic field along the $x$ axis. (c) Linear magnetic susceptibility of the CCW mode to the in-plane magnetic field.}
        \label{mag_excite}
\end{figure}

In this subsection, we show the magnetic susceptibility of the symmetry-adapted bases of the localized spins to corroborate the collective spin excitations in light of the frequency dependence.
We calculate the magnetic susceptibility $\chi_{\alpha \mu}(\omega)$ written in the symmetry-adapted basis as 
\begin{align}
    \Delta \tilde{\psi}_{\alpha}(\omega)=\chi_{\alpha \mu}(\omega)B_{\mu}(\omega),
\end{align}
where we defined the Fourier component of $\Delta\tilde{\psi}_{\alpha}(t)=\tilde{\psi}_{\alpha}(t)-\tilde{\psi}_{\alpha}(0)$, representing the modulation of localized spins in the symmetry-adapted representation.
The response function $\chi_{\alpha\mu}(\omega)$ is obtained by taking the Gaussian magnetic field $\bm{F}(t) =\bm{B}(t)$ [\Eqref{Gauss}], which is coupled to the localized spins.
The perturbation is expressed by the Hamiltonian
\begin{equation}
    H_\text{ext}(t) = - \sum_{a} \bm{B}(t) \cdot \bm{\mathrm{S}}_a .
\end{equation}
The perturbation takes place without breaking the translation symmetry, and the time-evolution can be traced in parallel between different momentum as in the case of the uniform vector potential [\Eqref{vonNeumann}].

We plot the magnetic susceptibility of the symmetry-adapted basis belonging to the different symmetry in \figref{mag_excite}.
In \figref{mag_excite} (a), the magnetic susceptibility $\chi_{\alpha z}$ with $\alpha = \mathrm{A}_1^{1z},\mathrm{A}_1^{3z},\mathrm{A}_1^{3xy},\mathrm{A}_2^{3xy}$ (AS modes) are shown.
Consistent with the symmetry analysis in Sec.~\ref{selection-rule-response}, the out-of-plane magnetic field is coupled to the AS modes but makes no contribution to the CW and CCW modes.
On the other hand, the CW and CCW modes respond to the in-plane magnetic fields as observed in the plots of $\chi_{\alpha x}$ of \figref{mag_excite} (b,c).

As a result, the spectrum of magnetic susceptibility is determined by three types of collective modes and consists of four peaks, that is one for the AS modes, two for the CW, and one for the CCW.
The number of the collective modes is consistent with the analysis of magnon spectrum~\cite{Akagi2013}, and the lowest-energy ($\omega \simeq 0.09$) peak comes from the magnon band whose gap energy at $\bm{k}=\bm{0}$ is due to the magnetic anisotropy $K_z$ in \Eqref{Ham}.
We note that only the $\bm{k} = \bm{0}$ spin excitation is present since the system is perturbed by the uniform light field parametrized by $\bm{E} (t)$.

\subsection{Optical Conductivity}\label{opt}

In this subsection, we investigate the linear optical conductivity.
The linear optical conductivity is written by
\begin{align}
    j_{\mu}(\omega)=\sigma_{\mu\nu}(\omega)E_{\nu}(\omega),
\end{align}
where $\mu,\nu = x,y$ since the electronic Hamiltonian is built onto the two-dimensional system.
$j_{\mu}(\omega)$ is the Fourier component of the electric current of \Eqref{electric-current}.

\begin{figure}
        \centering
        \includegraphics[width=\linewidth]{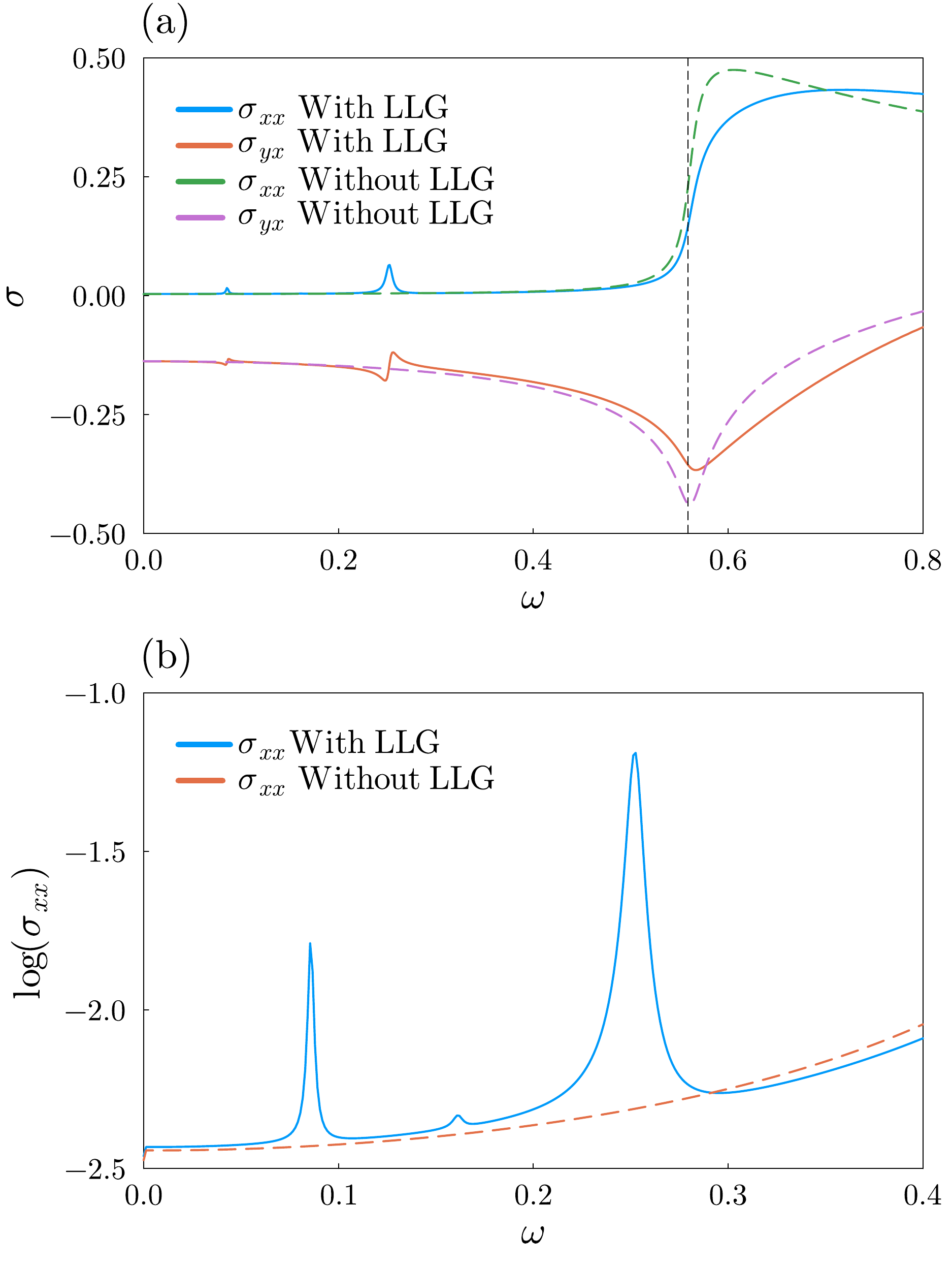}
        \caption{(a) Linear optical conductivity of the system. The blue
        solid line and the green dashed line indicate the longitudinal optical conductivity
        with and without updating the spin configurations, respectively. The orange
        solid line and the purple dashed line indicate the transverse optical conductivity
        with and without updating the spin configurations, respectively. The black dashed line indicates the band gap frequency. (b) Longitudinal optical conductivity at log scale. The blue
        solid line and the orange dashed line indicate the calculation
        with and without updating the spin configurations.}
        \label{optical_cond}
\end{figure}

We plot the frequency dependence of the linear optical conductivity $\sigma_{\mu\nu}(\omega)$ in \figref{optical_cond}.
The optical gap is at $\omega \simeq 0.55$ [vertical dashed line in \figref{optical_cond}(a)], indicating the insulating state of the adopted model.
In \figref{optical_cond}(a), we show the longitudinal optical conductivity $\sigma_{xx}$ and the transverse optical conductivity $\sigma_{yx}$.
The solid lines (`With LLG') represent the optical conductivity incorporating the effect of the localized spin dynamics following the LLG equation, while the dashed lines (`Without LLG') represent the optical conductivity without the effect.
In \figref{optical_cond}(b), we show the longitudinal optical conductivity $\sigma_{xx}$ with and without the LLG simulation in log scale.

There are two consequences of spin dynamics on the optical conductivity.
First, $\operatorname{Re}\sigma(\omega)$ with the LLG simulation shows the three resonance peaks in the in-gap regime, which are absent in the independent particle approximation.
Due to the increased degrees of freedom of the local spins, more optical conductivity peaks are observed in the in-gap regime than in the studies \cite{Iguchi,Kaneko2021}.
The peaks at $\omega\simeq0.09, 0.25$ correspond to CW modes as inferred from the magnetic susceptibility [\figref{mag_excite}~(b)], and the peak at $\omega\simeq0.16$ is for CCW modes [see \figref{mag_excite}(c)].
The peak height for $\omega\simeq0.16$ is much smaller than that at $\omega\simeq0.09,0.25$.
The prominent difference can be understood by the electromagnetic susceptibility corroborated in Sec.~\ref{ele}.

Second, the spin dynamics leads to the modification of the optical conductivity in the spectrum above the band gap as well as the in-gap optical excitations.
This modification is also carried by the collective spin dynamics, while the localized spins are driven by the electronic spin excitations responding to the irradiating light.
It is noteworthy that the in-gap and above-gap optical excitations are in the trade-off relationship due to the sum rule $\int\operatorname{Re}\sigma_{\mu\nu}(\omega)d\omega = \mathrm{const.}$
In the spin-charge coupled system with the AIAO spin texture, the collective modes belonging to the different symmetry affect the optical conductivity. Therefore, we can selectively excite the collective modes which influence electronic transport.

\subsection{Electromagnetic Susceptibility}\label{ele}

\begin{figure}
        \centering
        \includegraphics[width=\linewidth]{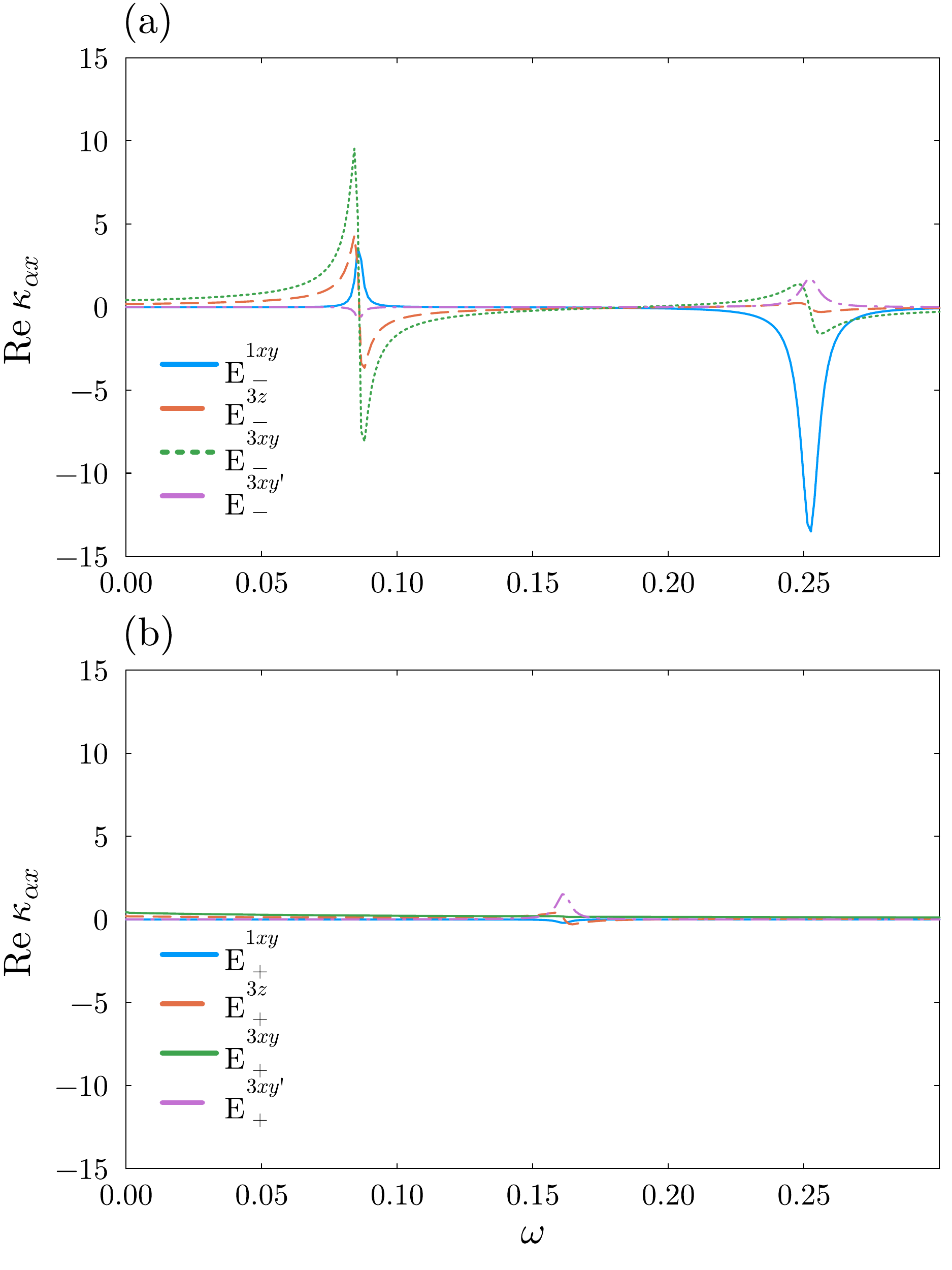}
        \caption{
        Linear electromagnetic susceptibility $\kappa_{\alpha x}$.
        The symmetry-adapted basis $\psi_\alpha$ is taken from (a) the CW modes and from (b) the CCW modes.
        }
        \label{ele_excite}
\end{figure}

Our focus is on the richer collective spin excitations originating from the AIAO state and on their impact on the electric response, and thus let us make more detailed investigations of the in-gap optical excitations by considering the electromagnetic susceptibility.
The electromagnetic susceptibility $\kappa_{\alpha\mu}(\omega)$ of the symmetry-adapted form of localized spin excitation $\psi_{\alpha}$ as
\begin{align}
    \Delta \tilde{\psi}_{\alpha}(\omega)=\kappa_{\alpha\mu}(\omega)E_{\mu}(\omega).
    \label{electromagnetic-susceptibility}
\end{align}
We plot the spectrum of the electromagnetic susceptibility $\kappa_{\alpha x}$ with the electric field $\bm{E} \parallel \hat{x}$ in \figref{ele_excite}.
Owing to the selection rule tabulated in \tabref{selection-rule-response}, only the CW and CCW modes contribute to the electromagnetic susceptibility $\kappa_{\alpha x}$.
Then, we show the electromagnetic susceptibility relevant to the CW modes [\figref{ele_excite}(a)] and CCW modes [\figref{ele_excite}(b)].
The electromagnetic susceptibility of the CCW mode is much smaller than that of the CW mode.

The in-gap peaks of the optical conductivity are caused by resonant excitation of collective spin dynamics, and thus peak amplitude implies the electrical activity of each collective excitation.
For instance, the electrical activity of the CCW mode is small as observed in the electromagnetic susceptibility plotted in \figref{ele_excite}(b), leading to the negligibly small peak in the optical conductivity spectrum [\figref{optical_cond}(b)].
More specifically, the mechanism of electric-active collective spins dynamics is as follows: first, the electric field coupled to the electron system stimulates the magnetization response of electrons' spins through the spin-orbit coupling [third term in \Eqref{Ham}].
Then, the perturbed spin moment of electrons induces the dynamics of the localized spins as a consequence of the exchange interaction between them [second term in \Eqref{Ham}].
The resultant coupling between the electric field and the localized spins indicates the electromagnetic response denoted by \Eqref{electromagnetic-susceptibility}.

The mechanism can be formulated by decomposing the electromagnetic susceptibility into two parts as  
\begin{align}
    \label{contribution}
    \kappa_{\alpha x}(\omega)
        &=\sum_{\beta}\chi^{\bm{\mathrm{S}}}_{\alpha\beta}(\omega)\kappa^{\bm{\sigma}}_{\beta x}(\omega),\\
        &\equiv \sum_{\beta} \kappa_{\alpha \beta x} (\omega).
        \label{electromagnetic_susceptibility_decomposition}
\end{align}
$\kappa_{\beta x}^{\bm{\sigma}}(\omega)$ is the bare electromagnetic susceptibility indicating the correlation between the electric field as 
\begin{equation}
    \sum_\mu c_a^\mu (\alpha) \Delta\Braket{\sigma_a^\mu} (\omega) = \kappa_{\alpha \nu}^{\bm{\sigma}} (\omega) E_\nu (\omega), 
\end{equation}
where we define the Fourier component of $\Delta\Braket{\sigma_a^\mu}(t)=\Braket{\sigma_a^\mu}(t)-\Braket{\sigma_a^\mu}(0)$, representing the modulation of itinerant spins in the symmetry-adapted representation.
This susceptibility does not include the dynamical effects of the localized spins, so we calculate $\kappa_{\alpha x}^{\bm{\sigma}}(\omega)$ by the real-time simulation without updating the local spin moments.

We also define $\chi_{\alpha\beta}^{\bm{\mathrm{S}}}(\omega)$, that is the magnetic susceptibility denoting the correlation between the symmetry-adapted modes ($\psi_\alpha$, $\psi_\beta$).
Based on the symmetry, $\chi_{\alpha\beta}^{\bm{\mathrm{S}}}(\omega)$ is finite when the symmetry-adapted bases $\alpha,\beta$ belong to the same mode.
The response function $\chi_{\alpha\beta}^{\bm{\mathrm{S}}}(\omega)$ is obtained by taking the Gaussian magnetic field, which is proportional to the symmetry-adapted basis $\psi_{\beta}$.
The perturbation is expressed by the Hamiltonian
\begin{equation}
    H_{\mathrm{ext}}^{\beta}(t)=-J\sum_{a\mu}\frac{B_0}{\sqrt{2\pi\sigma^2}}\mathrm{exp}\left(-\frac{(t-t_0)^2}{2\sigma^2}\right)c_{a}^{\mu}(\beta)\mathrm{S}_{a}^{\mu},
\end{equation}
where $B_0$ is the amplitude of the Gaussian magnetic field.

Combining $\chi_{\alpha\beta}^{\bm{\mathrm{S}}}(\omega)$ and $\kappa_{\beta x}^{\bm{\sigma}}(\omega)$, we can calculate the decomposed component $\kappa_{\alpha\beta x}(\omega)$.
As a result, the electromagnetic susceptibility $\kappa_{\alpha\mu}(\omega)$ of the collective mode $\psi_\alpha$ is determined by the interference of $\kappa_{\alpha\beta \mu}(\omega)$ with respect to the mediating modes labeled by $\beta$.

\begin{figure}
        \centering
        \includegraphics[width=\linewidth]{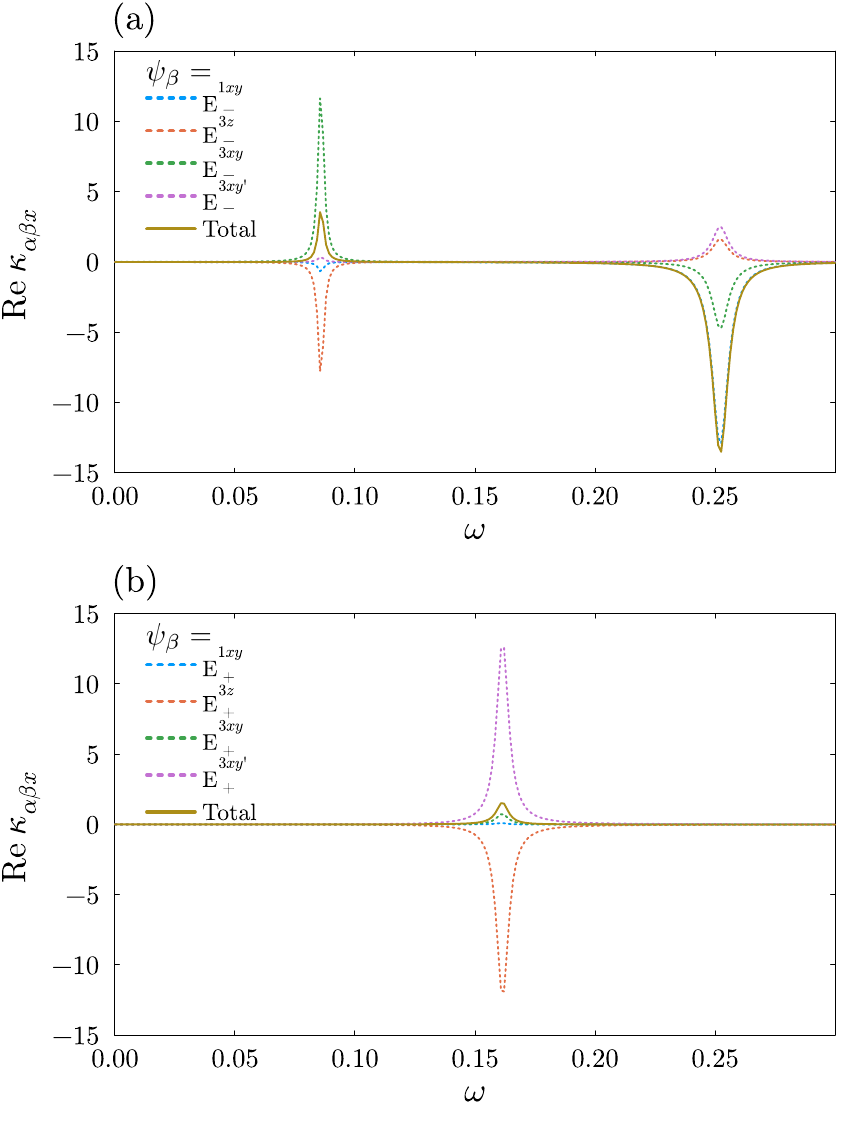}
        \caption{(a) The decomposition of the electromagnetic susceptibility of the collective mode $\psi_{\alpha}=\mathrm{E}_{-}^{1xy}$. (b) The decomposition of the electromagnetic susceptibility of the collective mode $\psi_{\alpha}=\mathrm{E}_{+}^{3xy'}$.}
        \label{decomposition}
\end{figure}

Taking the CW $\mathrm{E}_-^{1xy}$ and CCW modes $\mathrm{E}_+^{3xy'}$ for $\psi_\alpha$ in \Eqref{electromagnetic_susceptibility_decomposition}, we plot the decomposed electromagnetic susceptibility in \figref{decomposition}.
In the case of $\psi_\alpha = \mathrm{E}_{-}^{1xy}$ for example, the localized-spin response to the electric field is mediated by the fluctuations denoted by $\sum_\mu c_a^\mu (\beta) \Braket{\sigma_a^\mu}$ which are symmetry-adapted to the CW modes ($\mathrm{E}_{-}^{1xy}$, $\mathrm{E}_{-}^{3z}$, $\mathrm{E}_{-}^{3xy}$, and $\mathrm{E}_{-}^{3xy'}$).
Similarly, the electromagnetic susceptibility of the collective mode $\mathrm{E}_{+}^{3xy'}$ is mediated by those sharing the same symmetry with the CCW modes of localized spins.

As for the electromagnetic susceptibility of the CW mode, the interference of the mediating spin fluctuations results in a sizable electromagnetic susceptibility for $\psi_\alpha = \mathrm{E}_{-}^{1xy}$.
On the other hand, while each $\kappa_{\alpha\beta x}$ makes a significant contribution, destructive interference is observed in the case of the CCW mode, giving rise to a much smaller electromagnetic susceptibility in total.
These results imply that the CCW modes are less electrically active when compared to the CW modes and hence offer negligible optical conductivity [peak around $\omega \simeq 0.16$ in \figref{optical_cond}(b)].
The constructive and destructive interference effects are features unique to the system hosting the complex spin texture, and their identification has been accomplished by the developed efficient computational method for the real-time evolution of the spin-charge coupled system.

\section{Summary}
\label{summary}
In this study, we developed an efficient real-time simulation of the spin-charge coupled system and applied this method to the two-dimensional system with a complex spin texture.
We investigated how the collective excitation of the localized spins influences the optical response of the electronic system.
First, we identified the symmetry-adapted bases of the localized spins and further classified them with the circular polarization. 
This allowed us to clarify the activity of the collective modes of the localized spins to external stimuli.
Next, we calculated the linear response functions of the spin-charge coupled system.
The magnetic susceptibility is conveniently decomposed by the symmetry-adapted bases and is determined by the resonant dynamics of the localized spins when the frequency of magnetic fields is in the in-gap regime.
The observed collective modes have features similar to those of the skyrmionic system, implying the AIAO state is the minimum unit of the magnetic skyrmion.

The impact of the real-time simulation has been demonstrated by the calculations of the linear optical conductivity including the effect of the localized-spins dynamics.
The electrical activity of the collective spin motion results in the modification of the spectrum of optical conductivity such as in-gap peak structures.
The collective modes differ in their influence on the optical conductivity; the rotatory modes of the localized spins affect the optical conductivity, and the AS mode of the localized spins does not affect the optical conductivity in the two-dimensional system.
Finally, we discussed electromagnetic susceptibility, paying special attention to spin dynamics induced by the light whose frequency is below the band gap.
The electromagnetic susceptibilities for the CW and CCW modes are elucidated by the decomposition into the bare electromagnetic susceptibility and magnetic susceptibility.
As a result, the interference effect plays a key role in determining the contrasting electrical activity of the CW and CCW modes.

To conclude, we systematically investigated the spin-charge coupling by real-time simulations of the system with the AIAO magnetic order, which is considered the minimal unit of magnetic skyrmion.
The diverse collective excitations arising from the magnetic structure lead to multiple peak structures in the optical conductivity spectrum and exhibit different electrical activities due to interference effects between spin fluctuations.
The essential ingredient is the optically active collective spin dynamics in the gapped system.
Therefore, searching for a real material should be in the realm of inversion broken semiconductors or insulators with optically active magnons, such as $\mathrm{GaV_4S_8}$ \cite{Padmanabhan2019, Kézsmárki2015} and $\mathrm{Cu_2OSeO}$ \cite{Seki2012,White2014}.

\section*{acknowledgement}
This work is supported by Grant-in-Aid for Scientific Research from JSPS, KAKENHI Grant No.~JP23K13058 (H.W.), No.~22H00290(T.N.), No.~24K00581(T.N.), No.~21H04990 (R.A.), No.~JP21H04437 (R.A.), No.~19H05825 (R.A.), JST-PRESTO No.~JPMJPR20L7 (T.N.), JST-CREST No.~JPMJCR18T3, No.~JPMJCR23O4(R.A.), JST-ASPIRE No.~JPMJAP2317 (R.A.), JST-Mirai No.~JPMJMI20A1 (R.A.). K.H. was supported by the Program for Leading Graduate Schools (MERIT-WINGS).

\clearpage

\twocolumngrid

\bibliography{refs}

\end{document}